\documentclass[aps,prd,preprintnumbers,showpacs,showkeys,nofootinbib,
superscriptaddress,fleqn,floatfix,tightenlines,10pt]{revtex4-1}
\usepackage{amsmath,amsfonts,amssymb,amscd,amsxtra,amsthm}
\usepackage{graphicx}  
\usepackage{epstopdf}
\usepackage{dcolumn}  
\usepackage{bm}          
\usepackage{slashed}
\usepackage{cancel}
\usepackage{float}
\usepackage{mathtools}
\usepackage{amsbsy}
\usepackage{amstext}

\usepackage[utf8]{inputenc} 
\usepackage{booktabs}
\usepackage[normalem]{ulem} 
\usepackage[dvipsnames]{xcolor} 
\usepackage{tabularx}
\usepackage{enumitem}
\usepackage{array}
\usepackage{slashed}
\usepackage{tikz}
\usepackage{float}
\usepackage{multirow}
\renewcommand\sout{\bgroup \color{red} \ULdepth=-.5ex \ULset}

\makeatletter

\begin{document}
\preprint{INHA-NTG-04/2018}
\title{$K^0\Lambda$ photoproduction off the neutron with nucleon 
resonances}
\author{Sang-Ho Kim}
\email[E-mail: ]{sangho.kim@apctp.org}
\affiliation{Asia Pacific Center for Theoretical Physics (APCTP), 
 Pohang 37673, Republic of Korea}
\author{Hyun-Chul Kim}
\email[E-mail: ]{hchkim@inha.ac.kr}
\affiliation{Department of Physics, Inha University, Incheon 22212,
 Republic of Korea}
\affiliation{Advanced Science Research Center, Japan Atomic Energy
  Agency, Shirakata, Tokai, Ibaraki, 319-1195, Japan} 
\affiliation{School of Physics, Korea Institute for Advanced Study 
 (KIAS), Seoul 02455, Republic of Korea}
\date{\today}
\begin{abstract}
We investigate kaon photoproduction off the neutron target,
i.e., $\gamma n \to K^0 \Lambda$, focusing on the role of nucleon
resonances given in the Review of Particle Data Group in the range  
of $\sqrt{s} \approx 1600 - 2200$ MeV. We employ an effective
Lagrangian method and a Regge approach. The strong couplings of  
nucleon resonances with $K\Lambda$ vertices are constrained by quark
model predictions. The numerical results of the total and differential 
cross sections are found to be in qualitative agreement with the
recent CLAS and FOREST experimental data. We discuss the effects of
the narrow nucleon resonance $N(1685,1/2^+)$ on both the total and
differential cross sections near the threshold energy. In addition, we
present the results of the beam asymmetry as a prediction. 
\end{abstract}
\maketitle

\textbf{1.} Kuznetsov et al. reported the measurement of the cross
sections for $\eta$ photoproduction off the neutron, which shows a
narrow bump structure near the center-of-mass (CM) energy $W=1.68$ 
GeV~\cite{Kuznetsov:2006kt}. In the $\gamma p\to \eta p$ reaction,
there is only a small dip structure at the same energy.
The LNS-KEK Collaboration~\cite{Miyahara:2007zz},
the CB-ELSA and TAPS Collaborations in Bonn~\cite{Jaegle:2008ux}, and
the A2 Collaboration in
Mainz~\cite{Werthmuller:2013rba,Werthmuller:2014thb,
  Witthauer:2017get} have confirmed this feature of $\eta$
photoproduction off the neutron. This phenomena is often called the  
\textit{neutron anomaly} in $\eta$ photoproduction. However, there
is no consensus in the interpretations on the narrow enhancement at
$W=1.68$ GeV. In fact, the narrow nucleon resonance around $1.68$ GeV
was predicted by the chiral quark-soliton
model~\cite{Diakonov:1997mm,Polyakov:2003dx,Kim:2005gz,Yang:2013tka,
  YangKim} in which the neutron anomaly was explained in terms of the
different values of the $N(1685)\to N\gamma$ transition magnetic moments.
The A2 measurement of the helicity-dependent $\gamma n\to \eta n$ cross 
sections favors the existence of a narrow $P_{11}$
resonance~\cite{Werthmuller:2013rba}. On the other hand,
Ref.~\cite{Anisovich:2015tla, Anisovich:2017xqg} disputed that such the 
narrow enhancement arises from the interference between 
$N(1535,1/2^-)$ and $N(1650,1/2^-)$, based on the Bonn-Gatchina
multi-channel partial-wave analysis. However,
Ref.~\cite{Kuznetsov:2017qmo} refuted it in favor 
of the narrow $P_{11}$ nucleon resonance. In this situation it is of
great importance to scrutinize the narrow structure around 1.68 GeV
and the related neutron anomaly in other processes such as
$K^0\Lambda$ photoproduction.

In the present Letter, we investigate the $K^0\Lambda$ photoproduction 
off the neutron, focussing on the effects of the narrow resonance 
structure around $1.68$ GeV, which appeared in the $\gamma n \to \eta n$ 
reaction.
Since the threshold energy of the $\gamma n \to K^0 \Lambda$ is $1.61$ 
GeV, $K^0\Lambda$ photoproduction can provide a possible clue in 
understanding the nature of the narrow nucleon resonance $N(1685,1/2^+)$. 
In this regard, the investigation on $K^0\Lambda$ photoproduction will 
shed light on the neutron anomaly yet from the different facet. 
While the theoretical investigations of $\gamma n \to \eta n$ reaction
have been carried out extensively in the
literature~\cite{Choi:2005ki,Fix:2007st, 
Doring:2009qr,Shklyar:2006xw,Arndt:2009nv}, that of $K^0\Lambda$ 
photoproduction is very limited~\cite{Mart:2011ez,Mart:2011ey,
Mart:2013fia,Mart:2017xtf}.
Recently, the FOREST Collaboration at the Research Center for Electron
Photon Science, Tohoku University~\cite{Tsuchikawa:2016ixc,
  Tsuchikawa:2017tqm}  
and the CLAS Collaboration at the Thomas Jefferson National Accelerator
Facility~\cite{Compton:2017xkt} have announced the experimental data
on the total and differential cross sections of $K^0 \Lambda$
photoproduction off the neutron~\footnote{However, one should keep in
mind that both the experimental data from the CLAS and FOREST
Collaborations were taken from the deuteron target, certain effects
from the Fermi motion are involved in the course of extracting the
two-body experimental data.}.
Very recently, the beam-target helicity asymmetry $E$ is also measured
at the CLAS Collaboration~\cite{Ho:2018riy}.  
Thus, it is of great interest to examine theoretically the role of the
narrow nucleon resonance $N(1685,1/2^+)$ also in this $\gamma n \to K^0
\Lambda$ reaction. We will employ an effective Lagrangian approach in
which we can consider directly the nucleon resonances in the $s$
channel. We will introduce sixteen different nucleon resonances up to
2.2 GeV. In addition, we take into account the narrow nucleon
resonance $N(1685,1/2^+)$ corresponding to the narrow enhancement
found in $\eta$ photoproduction off the neutron. We also include the
$K^*$ Reggeon exchange in the $t$ channel, since it explains properly
the high-energy behavior of the total cross section. 

\vspace{1cm}

\begin{figure}[htp]
\centering
\includegraphics[width=11cm]{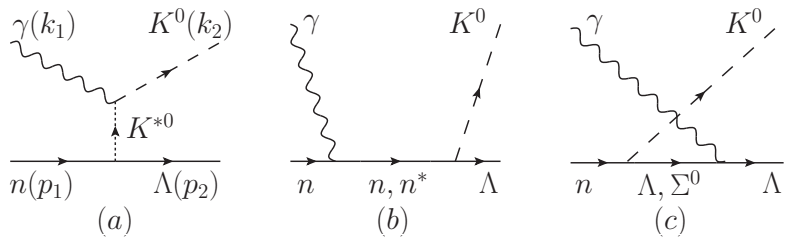}
\caption{Feynman diagrams for the $\gamma n \to K^0 \Lambda$ reaction.}
\label{fig:1}
\end{figure}

\textbf{2.} In an effective Lagrangian approach, the $\gamma n\to
K^0 \Lambda$ reaction can be represented by the 
tree-level Feynman diagram illustrated in Fig.~\ref{fig:1}. 
The notations of the four momenta of the incoming and outgoing
particles are given in Fig.~\ref{fig:1}(a) in which the $t$-channel
$K^*$ Reggeon exchange is depicted.
Other exchanges such as $K_1(1270,1^+),\, K_1(1400,1^+)$, and higher
strange mesons are excluded in the present process because of their
small photocouplings to the $K^0$ meson, e.g.,
$\mathrm{Br}(K^*(1410,1^-) \to K^0\gamma) < 2.2
\times10^{-4}$~\cite{Tanabashi:2018:abc}.

The $s$-channel diagrams shown in Fig.~\ref{fig:1}(b) include
contributions from the neutron and their resonances, generically. We
will consider the sixteen different nucleon resonances taken from the
Particle Data Group (PDG) data~\cite{Tanabashi:2018:abc}. 
On top of them, we include the narrow resonance $N(1685,1/2^+)$, which
corresponds to the narrow enhancement found in $\eta$
photoproduction~\cite{Kuznetsov:2006kt,Miyahara:2007zz,Jaegle:2008ux,
Werthmuller:2013rba, Werthmuller:2014thb,Witthauer:2017get}. 
$\Lambda$ and $\Sigma$ exchanges are included in the $u$-channel diagrams 
drawn in Fig.~\ref{fig:1}(c).   

The general expressions of the electromagnetic (EM) interaction
Lagrangians can be written as   
\begin{align}
\mathcal L_{\gamma K K^*} &=
g_{\gamma K K^*}^0 \epsilon^{\mu\nu\alpha\beta}
\partial_\mu A_\nu
( \partial_\alpha \bar K_\beta^{*0} K^0 + 
\bar K^0 \partial_\alpha K_\beta^{*0} ) ,
\cr
\mathcal L_{\gamma NN} &=
- \bar N \left[ e_N \gamma_\mu - \frac{e\kappa_N}{2M_N}
\sigma_{\mu\nu}\partial^\nu \right] A^\mu N ,                           
\cr
\mathcal L_{\gamma \Lambda \Lambda} &=
\frac{e \kappa_\Lambda}{2M_N}
\bar \Lambda \sigma_{\mu\nu} \partial^\nu A^\mu \Lambda ,
\cr
\mathcal L_{\gamma \Sigma \Lambda} &=
\frac{e\mu_{\Sigma\Lambda}}{2M_N}
\bar \Sigma^0 \sigma_{\mu\nu} \partial^\nu A^\mu \Lambda + \mathrm{H.c.},
\label{eq:BornLag1}
\end{align}
where $A_\mu$, $K$, $K^*$, and $N$ designate the fields for the photon, 
pseudoscalar kaon, vector kaon, and nucleon, respectively. 
$\Lambda$ and $\Sigma$ denote respectively the fields for the
ground-state hyperons. $M_N$ and $e_N$ stand respectively for the mass
and electric charge of the nucleon, whereas $e$ denotes the unit
electric charge. Since the neutron is involved in the present work,
we need only the magnetic term in the $\gamma NN$ vertex. 

Concerning the values of the coupling constants, $g_{\gamma K K^*}^0$ is 
determined by the experimental data for the decay width $\Gamma(K^*
\to K \gamma)$, resulting in $- 0.388 \, 
\mathrm{GeV}^{-1}$~\cite{Tanabashi:2018:abc}.   
The sign of the coupling is fixed from the quark model.
The anomalous and transition magnetic moments of the baryons are given  
by the PDG~\cite{Tanabashi:2018:abc}
\begin{align}
\kappa_N =-1.91, \,\,\, \kappa_\Lambda=-0.61, \,\,\,
  \mu_{\Sigma\Lambda}=1.61. 
\label{CouplConst1}
\end{align}

The effective Lagrangians for the meson-nucleon-hyperon interactions are 
given by
\begin{align}
\mathcal L_{K^* N \Lambda} &=
-g_{K^* N \Lambda} \bar N \left[ \gamma_\mu \Lambda - 
\frac{\kappa_{K^* N \Lambda}}{M_N+M_\Lambda} \sigma_{\mu\nu} \Lambda
\partial^\nu \right] K^{*\mu} + \mathrm{H.c.},
\cr
\mathcal L_{K N Y} &=
\frac{g_{K N Y}}{M_N+M_Y} \bar N \gamma_\mu \gamma_5 Y  \partial^\mu K 
+ \mathrm{H.c.},
\label{eq:BornLag2}
\end{align}
where $Y$ represents generically the fields for the hyperons ($\Lambda$
or $\Sigma^0$). The strong coupling constants are taken from the average
values of the Nijmegen soft-core potential (NSC97)~\cite{Stoks:1999bz} 
\begin{align}
&g_{K^*N\Lambda} = -5.19, \,\,\, \kappa_{K^*N\Lambda} = 2.79, \,\,\, \cr
&g_{KN\Lambda} = -15.5, \,\,\,\,\, g_{KN\Sigma} = 4.70     .
\label{CouplConst2}
\end{align}
Note that although we use the pseudovector coupling for the latter one
in Eq.~(\ref{eq:BornLag2}), the numerical results of the present work
almost do not change when the pseudoscholar coupling is employed,
since the effects of nucleon and hyperon exchanges turn out to be
tiny. 

In general, the invariant amplitude for photoproduction can be written
by   
\begin{align}
\mathcal{M}_h = I_h \bar u_\Lambda \mathcal{M}^\mu_h \epsilon_\mu u_N,
\label{eq:AmpNotation}
\end{align}
where $\epsilon_\mu$ represents the polarization vector of the incident 
photon. $u_N$ and $u_\Lambda$ denote the Dirac spinors for the
incoming nucleon and the outgoing $\Lambda$, respectively.
The isospin factors are given by $I_{K^*} = I_N = I_\Lambda = 1$ and 
$I_{\Sigma} = -1$ in the present process.
The effective Lagrangians of Eqs.~(\ref{eq:BornLag1}) and 
(\ref{eq:BornLag2}) being considered, the individual amplitudes for
the Born term are obtained as follows:
\begin{align}
\mathcal M_{K^*}^\mu&=
\frac{g_{\gamma K K^*}^0 g_{K^* N \Lambda}}{t-M_{K^*}^2} 
\epsilon^{\mu\nu\alpha\beta}
\left[ \gamma_\nu - \frac{i\kappa_{K^* N \Lambda}}{M_N+M_\Lambda} 
q_t^\lambda \sigma_{\nu\lambda} \right]
k_{1\alpha} k_{2\beta} ,                       
\cr
\mathcal M_N^\mu&= \frac{e\kappa_N}{2M_N} \frac{g_{K N \Lambda}}{2M_N}
\frac{1}{s-M_N^2}
\gamma_\alpha \gamma_5 (\rlap{/}{q_s}+M_N) \sigma^{\mu\nu} k_{1\nu} k_2^\alpha ,
\cr
\mathcal M_\Lambda^\mu&=
\frac{e \kappa_\Lambda}{2M_N} \frac{g_{K N \Lambda}}{M_N+M_\Lambda}
\frac{1}{u-M_\Lambda^2}
\sigma^{\mu\nu} k_{1\nu} (\rlap{/}{q_u}+M_\Lambda)
\gamma_\alpha \gamma_5 k_2^\alpha ,
\cr
\mathcal M_\Sigma^\mu&=
\frac{e \mu_{\Sigma\Lambda}}{2M_N} \frac{g_{K N \Sigma}}{M_N+M_\Sigma}
\frac{1}{u-M_\Sigma^2}
\sigma^{\mu\nu} k_{1\nu} (\rlap{/}{q_u}+M_\Sigma)
\gamma_\alpha \gamma_5 k_2^\alpha ,
\label{eq:BornEachAmp}
\end{align}
where $q_{t,s,u}$ designate the four momenta of the exchanged particles, 
i.e., $q_t=k_2-k_1$, $q_s = k_1+p_1$, and $q_u = p_2-k_1$.

Considering the finite sizes of hadrons, we need to introduce a
form factor at each vertex. It is of course well known that certain
ambiguities arise from the selection of hadronic form
factors, in particular, when higher spin resonant baryons are
involved~\cite{Vrancx:2011qv,Kristiano:2017qjq}.  Bearing in mind that
most approaches based on effective Lagrangians inevitably contain
uncertainties related to types of the form factors chosen, we will use
the following generic type for the $s$ and $u$-channel background
diagrams  
\begin{align}
F_B(q^2) =
\left[ \frac{\Lambda_B^4}
{\Lambda_B^4+\left(q^2-M_B^2\right)^2} \right]^2,
\label{eq:FF1}
\end{align}
where $q^2$ denotes the squared momentum of $q_{s,u}$ and $M_B$ the  
mass of the corresponding exchanged baryon $B$, respectively.
The form factor given in Eq.~\eqref{eq:FF1} tames sufficiently
unphysically increasing cross sections as $W$ increases. However, 
the gaussian-type form factors, which will be disscussed after
Eq.~\eqref{eq:FF2}, are employed for higher-spin baryon resonances, 
because they control more efficiently the resonance contributions such
that the cross sections are regulated and the resonance structures
are revealed reasonably well.

Although we are mainly interested in the vicinity of the threshold
energy for $K^0\Lambda$ photoproduction, future experiments are
exppected to cover higher energy regions. Thus, we employ the
$t$-channel Regge trajectory for the $K^*$-meson exchange and follow 
Refs.~\cite{Donachie2002,Guidal:1997hy}. This can be done by replacing
the Feynman propagator in Eq.~(\ref{eq:BornEachAmp}) with the Regge
one as 
\begin{align}
\label{eq:REGGEPRO}
\frac{1}{t-M_{K^*}^2} \to P_{K^*}^{\mathrm{Regge}} (t) =
\left( \frac{s}{s_0} \right)^{\alpha(t)-1}
\frac{\pi\alpha'}{\sin[\pi\alpha(t)]}
\left\{ \begin{array}{c} 1 \\ e^{-i\pi\alpha(t)} \end{array} \right\}
\frac{1}{\Gamma[\alpha(t)]},       
\end{align}
where either a constant phase $(1)$ or a rotating one $( 
e^{-i\pi\alpha(t)})$ can be considered for the Regge phase. 
The $K^*$ Regge trajectory reads~\cite{Guidal:1997hy}
\begin{align}
\alpha (t) = \alpha_{K^*} (t)=  0.83 t + 0.25,
\label{eq:ReggeTraj}
\end{align}
and $\alpha' \equiv \partial \alpha(t)/\partial t$ denotes the slope  
parameter. The energy-scale parameter is chosen to be $s_0 =
1\,\mathrm{GeV^2}$ for simplicity.
Consequently, the entire Born amplitude is written as
\begin{align}
\mathcal M_{\mathrm{Born}}&=
\mathcal M_{K^*} (t-M_{K^*}^2) P_{K^*}^{\mathrm{Regge}} (t)
+\mathcal M^{mag}_n  \,F_n (s)
+\mathcal M_\Lambda \, F_\Lambda (u)
+\mathcal M_\Sigma \, F_\Sigma (u) .
\label{eq:BornAmp}
\end{align}
Unlike the charged kaon production, all the terms are manifestly
gauge-invariant, so we do not need to introduce any prescription 
for gauge invariance. 

We also introduce $N^*$ contributions in the $s$ channel.
Among the nucleon resonances listed in the PDG, we take into account
sixteen different nucleon resonances in the range of
$\sqrt{s} \approx (1600 - 2200)$ MeV~\cite{Tanabashi:2018:abc}, including
the narrow $N(1685,1/2^+)$ in addition.
We first express the effective Lagrangians for the EM transitions 
of the nucleon resonances 
\begin{align}
\mathcal{L}^{1/2^\pm}_{\gamma  N N^*} &= 
\frac{eh_1}{2M_N} \bar N \Gamma^\mp
\sigma_{\mu\nu} \partial^\nu A^\mu N^* + \mathrm{H.c.} ,               
\cr
\mathcal{L}^{3/2^\pm}_{\gamma N N^*}&= 
-ie \left[ \frac{h_1}{2M_N} \bar N \Gamma_\nu^\pm
 - \frac{ih_2}{(2M_N)^2} \partial_\nu \bar N
 \Gamma^\pm \right] F^{\mu\nu} N^*_\mu + \mathrm{H.c.},   
\cr
\mathcal{L}^{5/2^\pm}_{\gamma N N^*} &=
e\left[ \frac{h_{1}}{(2M_N)^2} \bar N \Gamma_\nu^\mp
-\frac{ih_{2}}{(2M_N)^3} \partial_\nu \bar N
\Gamma^\mp \right] \partial^\alpha F^{\mu\nu}
N^*_{\mu\alpha} + \mathrm{H.c.} ,
\cr
\mathcal{L}^{7/2^\pm}_{\gamma N N^*} &=
ie \left[ \frac{h_{1}}{(2M_N)^3} \bar N \Gamma_\nu^\pm
-\frac{ih_{2}}{(2M_N)^4} \partial_\nu \bar N
\Gamma^\pm \right] \partial^\alpha \partial^\beta F^{\mu\nu}
N^*_{\mu\alpha\beta} + \mathrm{H.c.} ,
\label{eq:ResLag1}
\end{align}
where the spin and parity are given in superscripts.
$N^*$, $N^*_\mu$, $N^*_{\mu\alpha}$, and $N^*_{\mu\alpha\beta}$ stand for the 
spin-1/2, -3/2, -5/2, and -7/2 nucleon-resonance fields, respectively,
with
\begin{align}
\Gamma^{\pm} = \left(
\begin{array}{c}
\gamma_5 \\ I_{4\times4}
\end{array} \right) ,
\,\,\,\,
\Gamma_\nu^{\pm} = \left(
\begin{array}{c}
\gamma_\nu \gamma_5 \\ \gamma_\nu
\end{array} \right) .
\label{eq:GammaPM}
\end{align}
$h_i$ designate the EM transition coupling constants and can be calculated
from the Breit-Wigner helicity amplitudes $A_i$ given in the PDG.
We refer to Refs.~\cite{Oh:2007jd,Oh:2011} for the explicit relations
between them. It is found that the values of the $A_i$ for the 2018
edition of Review of Particle Physics~\cite{Tanabashi:2018:abc} are
almost the same as those for the 2016 edition
~\cite{Patrignani:2016xqp} except for the $N(1650,1/2^-)$. It is
changed from $-50 \pm 20$ to $-10$
[$10^{-3}/\sqrt{\mathrm{GeV}}$]. We want to mention that we use the
data on $N(1650,1/2^-)$ taken from the previous edition whereas those
on excited nucleons are employed from the updated 2018 edition of PDG.  
All the relevant values are tabulated in Table~\ref{TAB1}, where we adopt 
the central values of $A_i$. The electromagnetic coupling of the narrow 
resonance $N(1685,1/2^+)$ is taken from Ref.~\cite{Yang:2010}. 
As for the full decay width, the resonances less than 1800 MeV have rather
small values ($\simeq 130$ MeV) compared to those of the higher ones that
give $200-400$ MeV~\cite{Tanabashi:2018:abc}. 
In the present numerical calculation, we use the values in parentheses in 
Table~\ref{TAB1}.

\begin{table}[h]
\caption{The sixteen nucleon resonances listed by the Particle Data Group 
(PDG)~\cite{Tanabashi:2018:abc} and information on their electromagnetic 
couplings. The helicity amplitudes $A_{1/2,\,3/2}$
[$10^{-3}/\sqrt{\mathrm{GeV}}$] are obtained from
Ref.~\cite{Tanabashi:2018:abc}.
In addition, we introduce the narrow nucleon resonance in the last row of 
this Table, which corresponds to the narrow enhancement in $\eta$
photoproduction~\cite{Kuznetsov:2006kt,Miyahara:2007zz,Jaegle:2008ux,
Werthmuller:2013rba,Werthmuller:2014thb,Witthauer:2017get}.}
\label{TAB1}
\begin{tabular}{c||cc|cccc}
\hline
State&Rating&Width [MeV]
&$A_{1/2}$&$A_{3/2}$&$h_1$ &$h_2$ \\
\hline
$N(1650,1/2^-)$&****&100-150(125)
&$-50 \pm 20$~\cite{Patrignani:2016xqp}&$\cdots$
&$-0.31$&$\cdots$ \\
$N(1675,5/2^-)$&****& 130-160(145)
&$-60 \pm 5 $&$-85 \pm 10$&$4.88$&$5.45$ \\
$N(1680,5/2^+)$&****& 100-135(120)
&$\approx 30$& $\approx -35$
&$-7.44$&$8.57$ \\
$N(1700,3/2^-)$&***&100-300(200)
&$25 \pm 10$&$-32 \pm 18$&$-1.43$&$1.64$ \\
$N(1710,1/2^+)$&****& 80-200(140)
&$-40 \pm 20$&$\cdots$&$0.24$&$\cdots$ \\
$N(1720,3/2^+)$&****&150-400(250)
&$-80\pm 50$&$-140 \pm 65$&$1.50$&$1.61$ \\
$N(1860,5/2^+)$&**&300
&$21 \pm 13$&$34 \pm 17$&$0.28$&$1.09$ \\
$N(1875,3/2^-)$&***& 120-250(200)
&$10 \pm 6$&$-20 \pm 15$&$-0.55$&$0.54$ \\
$N(1880,1/2^+)$&*** & 200-400(300)
&$-60 \pm 50$&$\cdots$&$0.31$&$\cdots$ \\
$N(1895,1/2^-)$&****& 80-200(120)
&$13 \pm 6$&$\cdots$&$0.067$&$\cdots$ \\
$N(1900,3/2^+)$&**** & 100-320(200)
&$0\pm 30$&$-60 \pm 45$&$0.29$&$-0.56$ \\
$N(1990,7/2^+)$&**&  100-320(200)
&$-45 \pm 20$&$-52 \pm 27$&$6.92$&$7.54$ \\
$N(2000,5/2^+)$&**&300
&$-18 \pm 12$&$-35 \pm 20$&$-0.47$&$-0.56$ \\
$N(2060,5/2^-)$&*** & 300-450(400)
&$25 \pm 11$&$-37 \pm 17$&$0.027$&$-2.87$ \\
$N(2120,3/2^-)$&*** & 260-360(300)
&$110 \pm 45$&$40 \pm 30$&$-1.71$&$2.41$ \\
$N(2190,7/2^-)$&****& 300-500(400)
&$-15 \pm 13$&$-34 \pm 22$&$-1.57$&$-0.62$ \\
\hline
$N(1685,1/2^+)$&   &30
&$ $&$ $&$-0.315$~\cite{Yang:2010}&$ $ \\
\hline
\end{tabular}
\end{table}

The effective Lagrangians for the strong interactions are written as 
\begin{align}
\mathcal{L}^{1/2^\pm}_{K \Lambda N^*}&= 
 - i g_{K \Lambda N^*} \bar K \bar \Lambda 
 \Gamma^\pm N^* + \mathrm{H.c.},                                   
\cr
\mathcal{L}^{3/2^\pm}_{K \Lambda N^*}&=
 \frac{g_{K \Lambda N^*}}{M_K} \partial^\mu \bar K \bar \Lambda
 \Gamma^\mp N^*_\mu + \mathrm{H.c.},
\cr
\mathcal{L}^{5/2^\pm}_{K \Lambda N^*} &=
 \frac{ig_{K \Lambda N^*}}{M_K^2} \partial^\mu \partial^\nu \bar K
 \bar \Lambda \Gamma^\pm N^*_{\mu\nu} + \mathrm{H.c.},
\cr
\mathcal{L}^{7/2^\pm}_{K \Lambda N^*} &=
- \frac{g_{K \Lambda N^*}}{M_K^3} \partial^\mu \partial^\nu \partial^\alpha
\bar K  \bar \Lambda \Gamma^\mp N^*_{\mu\nu\alpha} + \mathrm{H.c.}.
\label{eq:ResLag2}
\end{align}
The strong coupling constants, $g_{K\Lambda N^*}$, can be extracted from the
quark model predictions where the information about the decay amplitude 
for the $N^* \to K \Lambda$ decay is given~\cite{Capstick:1998uh}. 
They are related by the following relation~\cite{Kim:2017nxg}:
\begin{align}
\label{eq:DA}
\langle K(\vec{q})\,\Lambda(-\vec{q},m_f) | -i \mathcal{H}_\mathrm{int} |
N^*({\bf 0},m_j) \rangle
=4 \pi M_{N^*} \sqrt\frac{2}{|\vec{q}|} \sum_{\ell,m_\ell}
\langle \ell\, m_\ell\, {\textstyle\frac{1}{2}}\, m_f | j \,m_j \rangle 
Y_{\ell,m_\ell} ({\hat q}) G(\ell),
\end{align}
where $\langle\ell\,m_\ell\,\frac{1}{2}\,m_f|j \,m_j \rangle$ and 
$Y_{\ell,m_\ell} ({\hat q})$ are the Clebsch-Gordan coefficients and
spherical harmonics, respectively.
The decay width is then obtained from the partial-wave decay amplitude 
$G(\ell)$
\begin{align}
\label{eq:DA2}
\Gamma (N^* \to K \Lambda) = \sum_\ell |G(\ell)|^2 .
\end{align}
The spin and parity of the nucleon resonance impose constraints on the 
relative orbital angular momentum $\ell$ of the $K\Lambda$ final state.
In the case of a $j^P = \frac{1}{2}^-$ resonance, the relative orbital
angular momentum is restricted by the angular momentum conservation, so 
the $s$ wave ($\ell = 0$) is only possible.
Similarly, for the resonances of $j^P =(1/2^+,\,3/2^+)$, $j^P = 
(3/2^-,\,5/2^-)$, $j^P = (5/2^+,7/2^+)$, and $j^P =7/2^-$, the
final-particle states are in the relative $p$, $d$, $f$, and $g$ waves, 
respectively. As a result, the relations between the decay
amplitudes and the strong coupling constants for the decays of the 
$j^P = (1/2^\pm,\,3/2^\pm,\,5/2^\pm, 7/2^\pm)$ resonances into the final state
are derived as follows:
\begin{align}
G\left(\frac{1+P}{2}\right) &= \mp
\sqrt{\frac{|\vec{q}|(E_{\Lambda} \mp M_{\Lambda})}{4\pi M_{N^*}}} 
g_{K \Lambda N^*}\,\,\,\,\mathrm{for}\,\,\,\,N^*(1/2^P),            
\cr
G\left(\frac{3-P}{2}\right) &= \pm
\sqrt{\frac{|\vec{q}|^3(E_{\Lambda} \pm M_{\Lambda})}{12\pi M_{N^*}}}
\frac{g_{K \Lambda N^*}}{M_K}\,\,\,\,\mathrm{for}\,\,\,\,N^*(3/2^P), 
\cr
G\left(\frac{5+P}{2}\right)  &= \mp
\sqrt{\frac{|\vec{q}|^5(E_{\Lambda} \mp M_{\Lambda})}{30\pi M_{N^*}}}
\frac{g_{K \Lambda N^*}}{M_K^2}\,\,\,\,\mathrm{for}\,\,\,\,N^*(5/2^P).
\cr
G\left(\frac{7-P}{2}\right)  &= \pm
\sqrt{\frac{|\vec{q}|^7(E_{\Lambda} \pm M_{\Lambda})}{70\pi M_{N^*}}}
\frac{g_{K \Lambda N^*}}{M_K^3}\,\,\,\,\mathrm{for}\,\,\,\,N^*(7/2^P),
\label{R12}
\end{align}
where the magnitude of the three-momentum and the energy for the
$\Lambda$ in the rest frame of the resonance are given respectively as 
\begin{align}
|\vec{q}|= \frac{1}{2M_{N^*}} 
\sqrt{[M_{N^*}^2 - (M_\Lambda + M_K)^2][M_{N^*}^2 - (M_\Lambda - M_K)^2]},
\,\,\,\,E_{\Lambda} = \sqrt{M_{\Lambda}^2 + |\vec{q}|^2}.
\label{eq:3-monen}
\end{align}

\begin{figure}[htp]
\centering
\includegraphics[width=6.5cm]{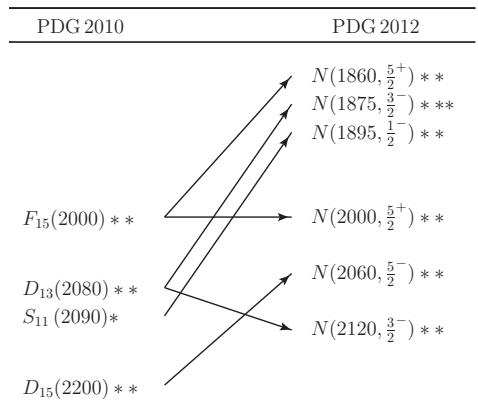}
\caption{Change of the $N^*$ spectrum from the 2010 edition of Review
  of Particle Physics to the 2012 edition.}
\label{fig:2}
\end{figure}

We should mention that the experimental data on the nucleon resonances in 
the 2012 edition of Review of Particle Physics~\cite{Beringer:1900zz} 
were much changed from those in the 2010 edition~\cite{Nakamura:2010zzi} 
(see Fig.~\ref{fig:2}).  
The $J^P = 5/2^+$ state $F_{15}(2000)$ is split into $N(1860,5/2^+)$ and 
$N(2000,5/2^+)$, whereas the $D_{13}(2080)$ breaks up into
$N(1875,3/2^-)$ and $N(2120,3/2^-)$. The $S_{11}(2090)$ is changed
into $N(1895,1/2^-)$ and the $N(2060,5/2^-)$ was previously identified
as $D_{15}(2200)$. Since the quark model predictions for the decay
amplitudes~\cite{Capstick:1998uh} are obtained from the resonances before 
the 2012 edition of Review of Particle Physics, we
thus make an assumption that the model values can be used for the
corresponding revised resonances. 

It is worthwhile to compare these coupling constants extracted from
the prediction of the quark model~\cite{Capstick:1998uh} with those
calculated from the experimental data on the branching 
ratios~\cite{Tanabashi:2018:abc}, although the signs of the couplings
can be fixed only in the quark models. In Table~\ref{TAB2}, we
summarize both values for the seventeen different nucleon resonances
under consideration. Only four resonances provide both of them. 
Comparing these two values, we find that they are close to each other.
Since only the experimental data exist for the $N(1880,1/2^+)$
and $N(1900,3/2^+)$, we determine the strong coupling constants for 
them by using the PDG data. Their signs are determined phenomenologically.
The last column in Table~\ref{TAB2} shows the couplings
$g_{K\Lambda N^*}$ that are finally determined.
They are mostly given within the range of the extracted coupling
constants from the quark model predictions~\cite{Capstick:1998uh} or
the PDG data~\cite{Tanabashi:2018:abc}.
Though we could reproduce the experimental data better by fitting the 
coupling constants, we have not performed it, because the main concern 
of the present work lies in understanding the role of each nucleon 
resonance and we want to avoid additional uncertainties arising from the
valuse of the strong coupling constants.

\begin{table}[h]
\caption{Information on the strong coupling constants of the nucleon
  resonances. The decay amplitudes $G(\ell)$ [$\sqrt{\mathrm{MeV}}$]
  are obtained from Ref.~\cite{Capstick:1998uh} and the branching
  ratios of $N^*$s to the $K\Lambda$ state are taken from
  Ref.~\cite{Tanabashi:2018:abc}.}
\label{TAB2}
\begin{tabular}{c||cc|cc|c}
\hline
State&$G(\ell)$
&$g_{K \Lambda N^*}$&$\Gamma_{N^* \to K\Lambda} / \Gamma_{N^*} [\%]$
&$|g_{K \Lambda N^*}|$&$g_{K \Lambda N^*}$(final) \\
\hline
$N(1650,1/2^-)$
&$-3.3 \pm 1.0$&$-0.78$&$5-15$
&$ 0.59-1.02$&$-0.78$ \\
$N(1675,5/2^-)$
&$0.4 \pm 0.3$&$1.23$& & &$1.23$ \\
$N(1680,5/2^+)$
&$\simeq 0.1 \pm 0.1$&$-2.84$& & &$-2.84$ \\
$N(1700,3/2^-)$
&$-0.4 \pm 0.3$&$2.34$& & & $2.34$ \\
$N(1710,1/2^+)$
&$4.7 \pm 3.7$&$-7.49$&$5-25$
&$ 4.2-9.4$&$-4.2$ \\
$N(1720,3/2^+)$
&$-3.2 \pm 1.8$&$-1.80$&$4-5$&$1.8-2.0$&$-1.1$ \\
$N(1860,5/2^+)$
&$-0.5 \pm 0.3$&$1.40$&seen& &$1.40$ \\
$N(1875,3/2^-)$
&$\simeq 1.7 \pm 1.0$&$-2.47$&seen& &$-2.47$ \\
$N(1880,1/2^+)$
&$$&$ $&$12-28$&$4.5-6.4$
&$3.0$ \\
$N(1895,1/2^-)$
&$2.3 \pm 2.7$&$0.34$&$13-23$
&$0.58-0.77$ &$0.34$ \\
$N(1900,3/2^+)$
&$$&$ $&$2-20$&$0.53-1.7$&$0.6$ \\
$N(1990,7/2^+)$
&$\simeq 1.5 \pm 2.4$&$0.61$& & &$0.61$ \\
$N(2000,5/2^+)$
&$-0.5 \pm 0.3$&$0.61$& & &$0.61$ \\
$N(2060,5/2^-)$
&$\simeq -2.2 \pm 1.0$&$-0.52$& seen& &$-0.52$ \\
$N(2120,3/2^-)$
&$\simeq 1.7 \pm 1.0$&$-1.05$& & &$-1.05$ \\
$N(2190,7/2^-)$
&$\simeq -1.1$&$0.67$& & &$0.67$ \\
\hline
$N(1685,1/2^+)$
&$ $&$ $& & &$-0.9$ \\
\hline
\end{tabular}
\end{table}

We can construct the individual amplitudes for the nucleon-resonance 
exchange using Eqs.~(\ref{eq:ResLag1}) and (\ref{eq:ResLag2}) in the
form of $\mathcal M = I_{N^*} \bar u_{\Lambda} \mathcal{M}_{N^*} u_N$
as in Eq.~(\ref{eq:AmpNotation}) with $I_{N^*} =1$:
\begin{align}
\mathcal M^{1/2^\pm}_{N^*}&=
\mp g_{K \Lambda N^*} \frac{eh_1}{2M_N}
 \frac{\Gamma^\pm(\rlap{/}{q_s} + M_{N^*})\Gamma^\mp}
{s-M_{N^*}^2+iM_{N^*}\Gamma_{N^*}}
 \sigma^{\mu\nu} k_{1\nu} \epsilon_\mu ,                      
\cr
\mathcal M^{3/2^\pm}_{N^*}&= 
i \frac{g_{K \Lambda N^*}}{M_K}
 \frac{\Gamma^\mp   k_2^\mu }
 {s-M_{N^*}^2+iM_{N^*}\Gamma_{N^*}}
\Delta_\mu^\rho (q_s)                          
\left[ \frac{eh_1}{2M_N} \Gamma_\lambda^\pm \mp 
       \frac{eh_2}{(2M_N)^2} \Gamma^\pm p_{1\lambda} \right]
(k_{1\rho} \epsilon^\lambda - k_1^\lambda \epsilon_\rho) ,
\cr
\mathcal M^{5/2^\pm}_{N^*}&= 
i \frac{g_{K \Lambda N^*}}{M_K^2}
\frac{\Gamma^\pm k_2^\mu k_2^\nu}
{s-M_{N^*}^2+iM_{N^*}\Gamma_{N^*}} 
\Delta_{\mu \nu}^{\rho \sigma} (q_s)                     
\left[ \frac{eh_1}{(2M_N)^2} \Gamma_\lambda^\mp \pm 
       \frac{eh_2}{(2M_N)^3} \Gamma^\mp p_{1\lambda} \right]
k_{1\sigma} (k_{1\rho} \epsilon^\lambda - k_1^\lambda \epsilon_{\rho}) ,
\cr
\mathcal M^{7/2^\pm}_{N^*}&= 
i \frac{g_{K \Lambda N^*}}{M_K^3}
\frac{\Gamma^\mp k_2^\mu k_2^\nu k_2^\alpha}
{s-M_{N^*}^2+iM_{N^*}\Gamma_{N^*}} 
\Delta_{\mu\nu\alpha}^{\rho\sigma\delta} (q_s)
\left[ \frac{eh_1}{(2M_N)^3} \Gamma_\lambda^\pm \pm 
       \frac{eh_2}{(2M_N)^4} \Gamma^\pm p_{1\lambda} \right]
k_{1\sigma} k_{1\delta} (k_{1\rho} \epsilon^\lambda - k_1^\lambda \epsilon_{\rho}) ,
\label{eq:ResEachAmp}
\end{align}
where $\Gamma_{N^*}$ designates the full decay width of $N^*$. 
The spin-$3/2$, -$5/2$, and -$7/2$ projection operators, given by 
$\Delta_\mu^\rho$, $\Delta_{\mu\nu}^{\rho\sigma}$, and 
$\Delta_{\mu\nu\alpha}^{\rho\sigma\delta}$, respectively, are represented in the 
Rarita-Schwinger formalism~\cite{Berends:1979rv,Behrends:1957rup,
Chang:1967zzc,Rushbrooke:1966zz} as in Refs.~\cite{Oh:2007jd,
Oh:2011,Kim:2011rm,Kim:2012pz}. The phase factors of the 
invariant amplitudes for the nucleon resonances cannot be determined
by symmetries only, so we regard them as free parameters. These
amplitudes are thus written by  
\begin{align}
\mathcal{M}_{\mathrm{Res}} =
\sum_{N^*} e^{i\psi_{N^*}} {\mathcal M_{N^*}} F_{N^*}(s),
\label{eq:ResAmp}
\end{align}
where the gaussian form factor is 
employed~\cite{Corthals:2005ce,DeCruz:2012bv}
\begin{align}
F_{\mathrm{N^*}}(q_s^2) = \mathrm{exp}
\left\{ - \frac{(q_s^2-M_{N^*}^2)^2}{\Lambda_{N^*}^4} \right\}.
\label{eq:FF2}
\end{align}

\begin{figure}[hb]
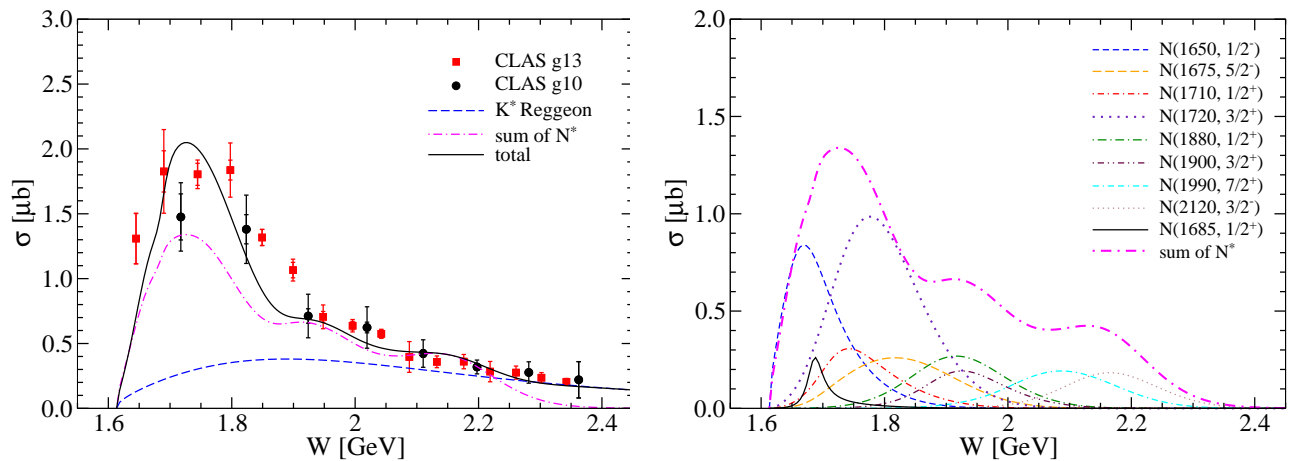

\includegraphics[width=8.3cm]{fig3a.eps} \,\,\,
\includegraphics[width=8.3cm]{fig3b.eps}
\caption{Left: Total cross section for the $\gamma n \to K^0 \Lambda$
  reaction as a function of the CM energy. The dashed (blue),
  dot-dashed (magenta), and solid (black) curves correspond to
  contribution from $K^*$ Reggeon exchange, that from the sum of
  $N^*$ exchanges, and the total contribution, respectively. The data
  are taken from the CLAS experiment~\cite{Compton:2017xkt}. Right:
  Each contribution to the $\gamma n \to K^0 \Lambda$ reaction for
  various nucleon resonances.}  
\label{fig:3}
\end{figure}
\textbf{3.} Before we present the numerical results, we need to
mention how the model parameters are fixed. 
The cutoff masses are fixed to be $\Lambda_{B(N,\Lambda,\Sigma),N^*}$
= 0.9 GeV for simplicity. We do not fit the values of the cutoff
masses to avoid additional uncertainties arising from them. We find
that at high energies above the CM energy $W = 2.2 \,\mathrm{GeV}$, 
where $K^*$ Reggeon exchange comes into a dominant play, the rotating 
Regge phase $(e^{-i\pi\alpha_{K^*}(t)})$ and the phase angle $\psi_{N^*} = \pi$ 
turn out to be the best choice.    

In the left panel of Fig.~\ref{fig:3}, the total cross section for the   
$\gamma n \to K^0 \Lambda$ reaction is drawn as a function of the
CM energy. The $N^*$ contributions are dominant in the lower-energy 
region ($W\lesssim 2.2$ GeV). $K^*$ Reggeon exchange in
the $t$ channel being included, the result is in agreement with the
CLAS data~\cite{Compton:2017xkt}. 
As $W$ increases, the $K^*$ Reggeon takes over $N^*$
contributions. Because of $K^*$ Reggeon exchange, the 
total cross section behaves asymptotically as $\sigma \sim 
s^{\alpha_{K^*}(0) -1}$ and describes the experimental data well. As
shown in the left panel of Fig.~\ref{fig:3}, the result is slightly
underestimated in the vicinity of the threshold energy, compared to
the CLAS data. Each contribution of various nucleon resonances is
drawn in the right panel of Fig.~\ref{fig:3}. The well-known
$N(1650,1/2^-)$ and $N(1720,3/2^+)$ are the most dominant
ones. While the
$N(1675,5/2^-)$, $N(1710,1/2^+)$, $N(1880,1/2^+)$, $N(1900,3/2^+)$,
$N(1990,7/2^+)$, and $N(2120,3/2^-)$
have sizable effects on the total cross section, 
all other resonances almost do not affect it, so we show only
the contributions of the nine nucleon resonances in the figure. 
Moreover, the $N(1685, 1/2^+)$ resonance has only a marginal effect
on the total cross section. 
Thus, as far as the results of the total
cross section are concerned, the present ones are more or less
similar to those of Ref.~\cite{Anisovich:2017afs} where the
Bonn-Gatchina coupled-channel partial-wave analysis was used.
In Ref.~\cite{Anisovich:2017afs}, it was shown that the partial waves
$J^P =1/2^\pm$ and $3/2^+$ contribute dominantly to the total cross
section and the narrow bump structure is not seen unlike the
$\gamma n \to \eta n$ cross section. However, the inclusion of the
$N(1685, 1/2^+)$ improves the data around $W=1.68$~GeV
in the present calculation.

\begin{figure}[hb]
\centering
\includegraphics[width=12cm]{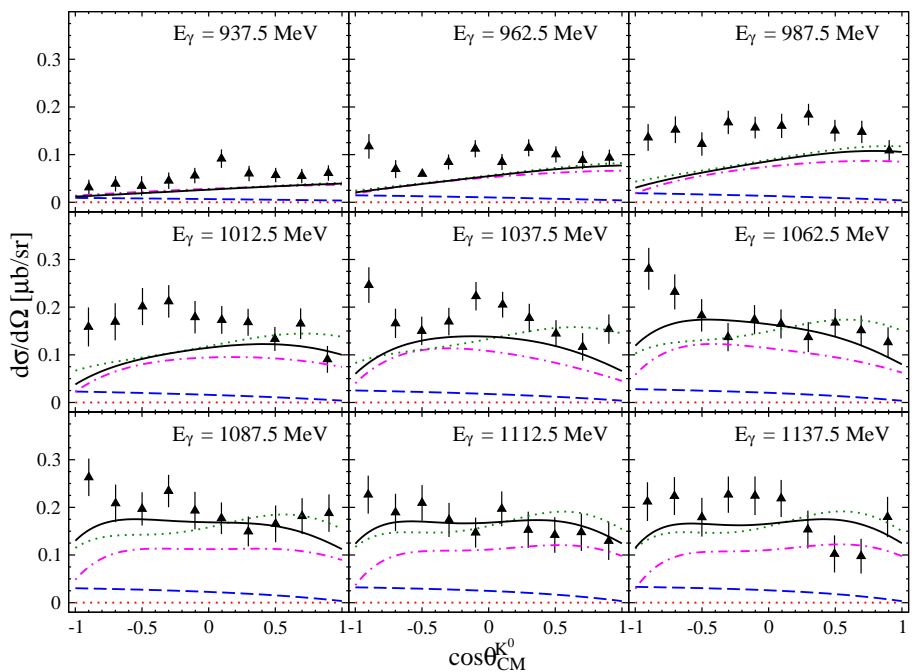} 
\caption{Differential cross section for the $\gamma n \to K^0
  \Lambda$ reaction as a function of $\cos\theta_{\mathrm{CM}}^{K^0}$
  for each beam energy. The dashed (blue), dot-dashed (magenta), and
  solid (black) curves correspond to the contribution from
  $K^*$ Reggeon exchange, that from the sum of $N^*$ exchanges, and
  the total contribution, respectively. The dotted (green) one
  indicates the total contribution without the effect of the narrow
  resonance $N(1685,1/2^+)$. The data are taken from the FOREST
experiment~\cite{Tsuchikawa:2017tqm}.} 
\label{fig:4}
\end{figure}
\begin{figure}[ht]
\centering
\includegraphics[width=15cm]{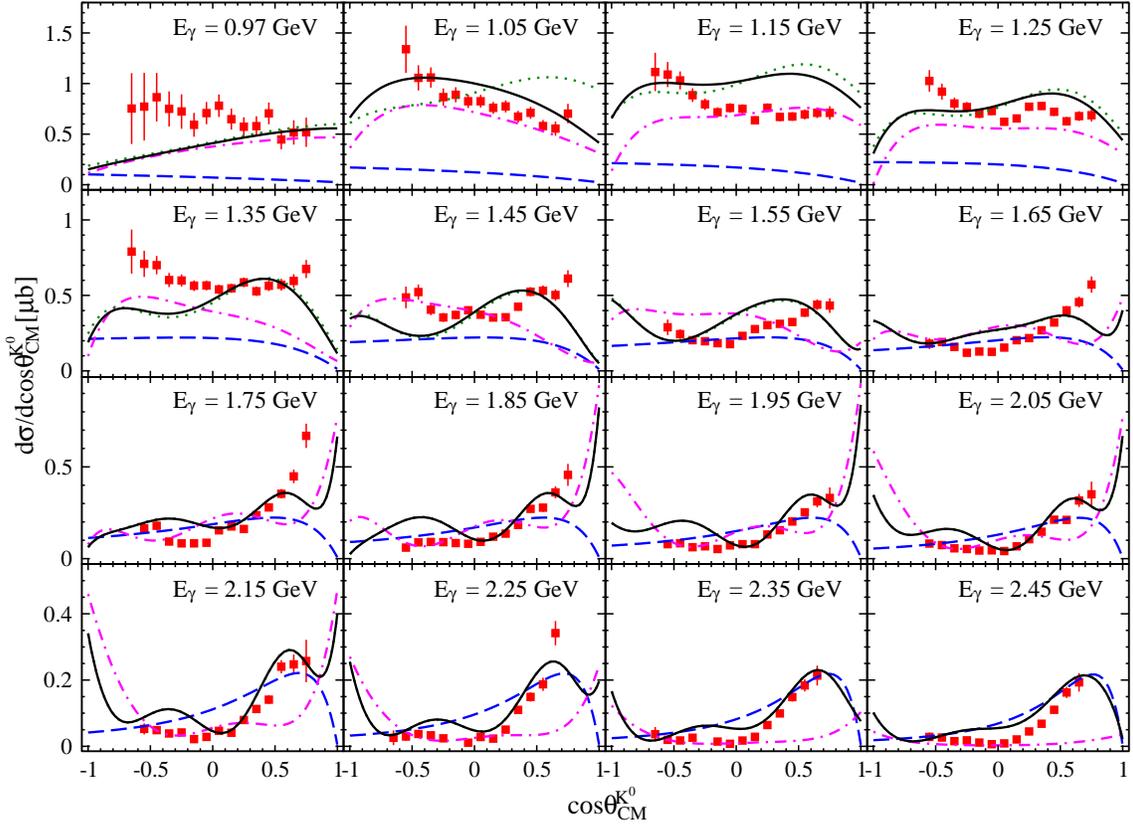} 
\caption{Differential cross section for the $\gamma n \to K^0
  \Lambda$ reaction as a function of $\cos\theta_{\mathrm{CM}}^{K^0}$
  for each beam energy. The notations are the same as in 
  Fig.~\ref{fig:4}. The data are taken from the CLAS
experiment~\cite{Compton:2017xkt}.} 
\label{fig:5}
\end{figure}
Figure~\ref{fig:4} draws the differential cross sections for the $\gamma n
\to K^0 \Lambda$ reaction as a function of 
$\cos\theta_{\mathrm{CM}}^{K^0}$, being compared with the FOREST 
experimental data~\cite{Tsuchikawa:2017tqm}. The photon energy is
varied from $E_\gamma=937.5$ MeV to $E_\gamma = 1137.5$ MeV. 
The dashed curve is drawn for the contribution of $K^*$ Reggeon
exchange. As expected, its effect is rather small in the range of the
photon energy given in Fig.~\ref{fig:4}. Here, main interest lies in
the effect of the narrow resonance $N(1685,1/2^+)$. While the dotted
curve is depicted without the $N(1685, 1/2^+)$ taken into account, the
solid one includes it. Though the effect of the $N(1685,1/2^+)$ is 
very small at smaller values of $E_\gamma$, it comes into play as
$E_\gamma$ increases. In particular, the experimental data of the
differential cross section at $E_\gamma=1037.5$ MeV and
$E_\gamma=1062.5$ MeV can be explained only by including the narrow
resonance $N(1685,1/2^+)$. Otherwise, the results would be
overestimated in the forward direction and would be underestimated in 
the backward direction. Although the $N(1685,1/2^+)$ does not give
any significant contribution to the total cross section, it is
essential to consider it to explain the differential cross section
data in the range of the photon energies $1037\,\mathrm{MeV} \le
E_\gamma \le 1062 \,\mathrm{MeV}$.

In Fig.~\ref{fig:5}, we compare the present results of the differential
cross section with the CLAS data~\cite{Compton:2017xkt}.
The CLAS experiment covers a much wider range of the photon energies
($0.97\,\mathrm{GeV}\le E_\gamma \le 2.45$ GeV) than the FOREST
experiment. The first three figures in the first row of
Fig.~\ref{fig:5} can be compared to the FOREST data given in
Fig.~\ref{fig:4}. Though there are some discrepancies between these
two experimental data, general tendency of the data is similar each
other. The present results are also in qualitative agreement with the
CLAS data. In particular, the narrow resonance $N(1685,1/2^+)$ pulls down
the differential cross section at $E_\gamma = 1.05$ GeV in
the forward direction. On the other hand, the $N(1685,1/2^+)$ makes
it enhanced in the backward direction, as we already discussed in
Fig.~\ref{fig:4}. As a result, the inclusion of the $N(1685,1/2^+)$
provides noticeably better agreement with the data. 
As $E_\gamma$ increases, i.e $E_\gamma \geq 1.8 
\,\mathrm{GeV}$ (or $W \geq 2.05 \,\mathrm{GeV}$), the results are in
good agreement with the CLAS data. This can be understood by $K^*$
Reggeon exchange which governs the $\gamma n \to K^0\Lambda$ process
in the higher energy region. 

We want to mention that we fit the value
of the $K\Lambda N(1685)$ coupling constant to be $g_{K \Lambda 
N(1685,1/2^+)} = -(0.8 - 1.1)$, which implies the branching ratio
Br($N(1685,1/2^+) \to K\Lambda$) = $(0.5 - 1.0)\,\%$ with
$\Gamma_{N(1685,1/2^+)} = 30$ MeV.
Consequently, we get the partial decay width 
$\Gamma_{N^*(1685) \to K\Lambda}$ to be $(0.15 - 0.30)$ MeV, whereas another
theoretical analysis based on the soliton picture yields $0.7\,(1.56)$ MeV
for $M_{N^*}$ = 1680(1730) MeV~\cite{Arndt:2003ga}.
We hope that future experiments may clarify these predictions.
Meanwhile, Ref.~\cite{Anisovich:2017afs} obtained the following upper
limit
\begin{align}
\sqrt{\mathrm{Br}(N(1685) \to K\Lambda)} \, A_{1/2}^n < 6 \times 10^{-3} \, 
\mathrm{GeV}^{-1/2} ,
\end{align}
which is consistent with our result, i.e., $(3.7 -5.2) \, \times 10^{-3} 
\,\mathrm{GeV}^{-1/2}$.

We have fixed the mass of this narrow resonance to be 
$M_{N^*}$ = 1685 MeV. On the other hand, a simultaneous analysis of
the $\gamma p \to K^+  \Lambda$ and $\gamma n \to K^0 \Lambda$
channels finds that the most appropriate mass is 1650
MeV~\cite{Mart:2011ey,Mart:2013fia}. 
The energies at $E_\gamma$ = 0.97 and 1.05 GeV in Fig.~\ref{fig:5} 
correspond approximately to the energies at $W$ = 1650 and 1685 MeV,
respectively. Thus, selecting such a low mass $M_{N^*}$ = 1650 MeV is
not suitable to describe the CLAS data in our calculation, since the
inclusion of the narrow resonance greatly improves the cross section
result at the energy $E_\gamma$ = 1.05 GeV.

\begin{figure}[ht]
\centering
\includegraphics[width=16cm]{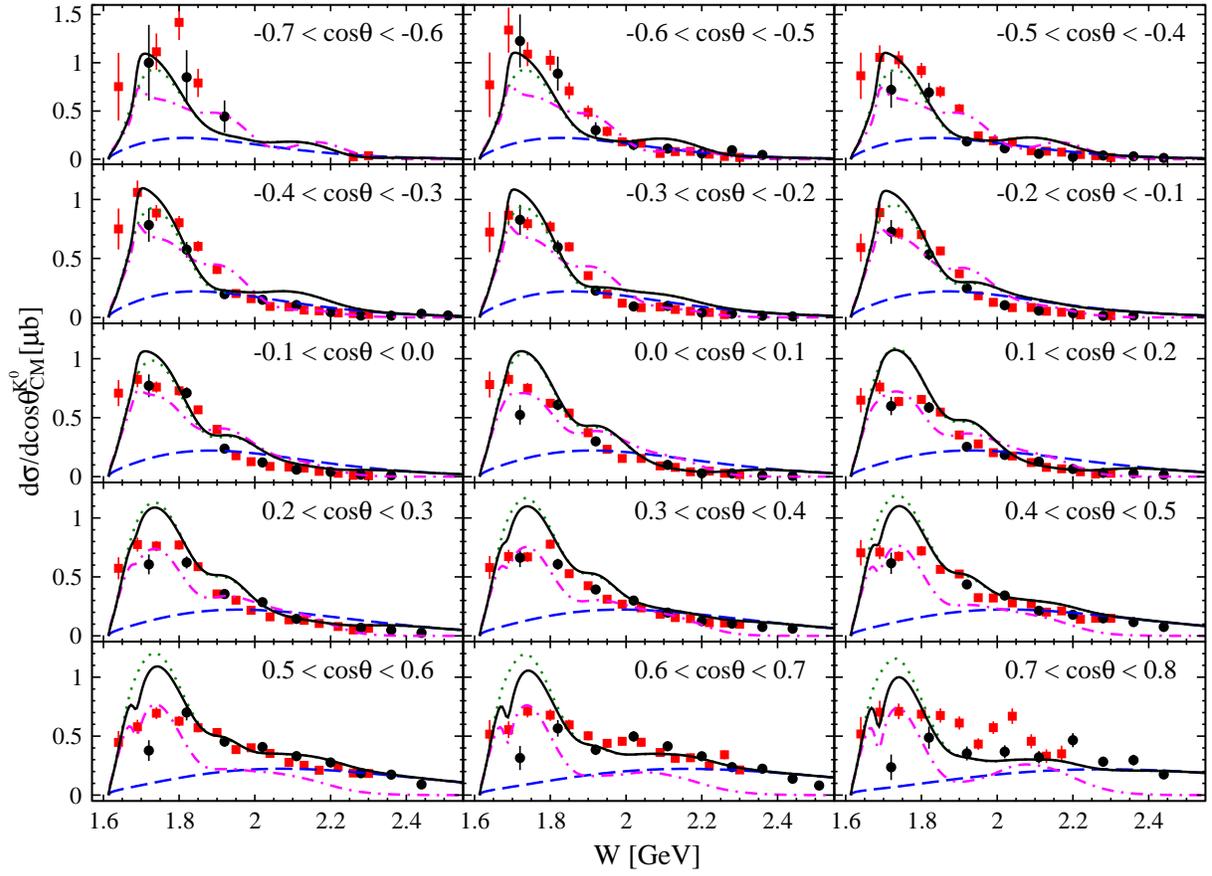} 
\caption{Differential cross section for $\gamma n \to K^0 \Lambda$
 as a function of $W$ for each $\cos\theta_{\mathrm{CM}}^{K^0}$.
The notations are the same as Fig.~\ref{fig:4}. 
The data are taken from the CLAS experiment~\cite{Compton:2017xkt}.} 
\label{fig:6}
\end{figure}

Figure~\ref{fig:6} draws the differential cross sections for the
$\gamma n \to K^0 \Lambda$ reaction as a function of the CM energy with
$\cos\theta_{\mathrm{CM}}^{K^0}$ fixed. As in the case of
Fig.~\ref{fig:5}, including the narrow resonance $N(1685,1/2^+)$
improves the description of the CLAS data near the threshold
region~\cite{Compton:2017xkt}. In particular, the $N(1685,1/2^+)$
enhances the differential cross section in the backward angle,
i.e. in the range of $-0.7 < \cos\theta_{\mathrm{CM}}^{K^0} < 0.0$ around
$W=1.68$ GeV.
On the other hand, the $N(1685,1/2^+)$ makes it reduced in the 
forward direction. Yet another noticeable feature is that, by the 
inclusion of the $N(1685,1/2^+)$, destructive effects between it and
other resonances begin to appear at the corresponding pole position as
the angle $\cos\theta$ increases as clearly seen in the last row of
Fig.~\ref{fig:6}.  This tendency is also shown in the CLAS data,
especially in the g10 ones. This can strongly support the evidence of
the existence of the narrow resonance $N(1685,1/2^+)$ in $K^0\Lambda$
photoproduction. One cannot explain these dip structures merely by
adjusting the model parameters of other resonances.

\begin{figure}[ht]
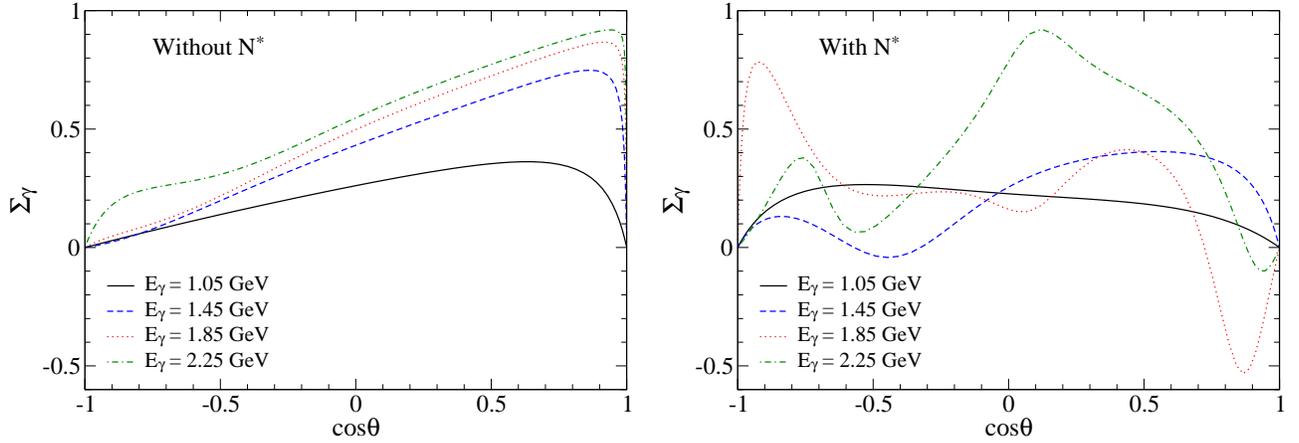

\centering
\includegraphics[width=8.3cm]{fig7a.eps} \,\,\,
\includegraphics[width=8.3cm]{fig7b.eps}
\caption{Beam asymmetry $\Sigma_{\vec\gamma n \to K^0 \Lambda}$ as a 
function of $\cos\theta$ for each beam energy when the nucleon resonances
are not included (Left) and included (Right), respectively.} 
\label{fig:7}
\end{figure}

It is of great importance to examine the beam asymmetry and other 
polarization observables of $K^0\Lambda$ photoproduction both 
experimentally and theoretically, since they can clarify more clearly
the role of nucleon resonances.
In Fig.~\ref{fig:7}, we draw the predictions of the beam asymmetry
$\Sigma_{\vec\gamma n \to K^0 \Lambda}$ as a function of $\cos\theta$ for
four different beam energies. 
The beam asymmetry is defined as follows: 
\begin{equation}
\label{eq:BA}
\Sigma_{\vec{\gamma} n \to K^0 \Lambda} = 
\frac{\frac{d\sigma}{d\Omega}_\perp-\frac{d\sigma}{d\Omega}_\parallel}
{\frac{d\sigma}{d\Omega}_\perp+\frac{d\sigma}{d\Omega}_\parallel},
\end{equation}
where the subscript $\perp$ means that the photon polarization vector
is perpendicular to the reaction plane whereas $\parallel$ denotes the
parallel photon polarization to it. The left panel of Fig.~\ref{fig:7}
depicts the results of the beam asymmetry without the nucleon
resonances. It shows that the beam asymmetry increases generally as
$\cos\theta$ increases. However, it falls off drastically in the very
forward direction. If the photon energy $E_\gamma$ increases, the
magnitudes of the beam asymmetry become larger. On the other hand,
when the nucleon resonances are included, the structure of the beam
asymmetry is entirely changed. At $E_\gamma=1.05$~GeV, the
$\cos\theta$ dependence of $\Sigma_\gamma$ is opposite to the case
without the nucleon resonances, that is, when $\cos\theta$ increases,
the beam asymmetry increases till around $\cos\theta=-0.3$ and then
slowly falls off. Hoever, if one increase the photon energy, the
results of $\Sigma_\gamma$ are changed dramatically. One can
understand this interesting feature of $\Sigma_\gamma$. Each
contribution of the nucleon resonances depends on $E_\gamma$. As
$E_\gamma$ increases, higher lying nucleon resonances comes into
play. It brings about the remarkable changes of the beam asymmetry. 
\vspace{1cm}

\textbf{4.} In the present work, we investigated $K^0\Lambda$
photoproduction, aiming at understanding the nature of 
the narrow resonance structure around $1.68$ GeV. We employed an 
effective Lagrangian method combined with a Regge approach. We
inlcuded seventeen different nucleon resonances in the $s$ channel
together with nucleon exchange as a background. In addition, we
considered $\Lambda$ and $\Sigma$ exchanges the $u$ channel. 
In the $t$ channel, we included $K^*$ Reggeon exchange which governs
the behavior of the $\gamma n\to K^0 \Lambda$ amplitude in higher
energy regions. Since charged kaon photoproduction has been widely
investigated in the literature, it is of great interest to compare the
role of each diagram in both the charged and neutral productions. 
Even though the photocouplings are different, the important
contributions in the $s$ channel are similar to each other.
For example, in Refs.~\cite{Shklyar:2005xg, Corthals:2005ce, 
Schumacher:2010qx, DeCruz:2012bv, Mart:2017mwj}, the $N(1650,1/2^-)$ and 
$N(1720,3/2^+)$ are the most significant ones and the $N(1900,3/2^+)$ is
also required for the description of the cross secton data. 

We have taken into account the nucleon resonances which appeared only
in the PDG data. 
We were able to reproduce the recent CLAS and FOREST data reasonably
well without any complicated fitting procedure, even though the nucleon 
resonances from the PDG data are considered only. 
The sign ambiguities of the strong coupling constants for the nucleon 
resonances were resolved by relating them to informaiton on the 
partial-wave decay amplitudes predicted by a quark model.
We found that the narrow nucleon resonance $N(1685,1/2^+)$ has a
certain contribution to the differential cross sections near the 
threshold energy. It enhances them in the backward direction while it
makes them decreased in the forward direction, whereas it is difficult
to see the effect of the $N(1685,1/2^+)$ on the total cross section.
It is interesting to extend our approach to the study of $K\Sigma$ 
photoproduction~\cite{Aguar-Bartolome:2013fqw} and $K\Lambda(K\Sigma)$ 
electroproduction~\cite{Carman:2012qj}.
Relevant works will appear elsewhere.

\begin{acknowledgments}
Authors are grateful to N. Compton for providing us with the 
CLAS experimental data. They want to express the gratitude to
T. Ishikawa, Y. Tsuchikawa, H. Shimizu, Y.~Oh, and Gh.-S. Yang for 
fruitful discussions. H.-Ch. K. is grateful to A. Hosaka, T. Maruyama,
M. Oka for useful discussions.  He wants to express his gratitude to
the members of the Advanced Science Research Center at Japan Atomic
Energy Agency for the hospitality, where part of the present work was
done. S.H.K. acknowledges support by the Ministry of Science, ICT $\&$  
Future Planning, Gyeongsangbuk-do and Pohang City.
This work was supported by the National Research Foundation of Korea
(NRF) grant funded by the Korea government(MSIT)
(No. 2018R1A5A1025563). 
\end{acknowledgments}


\end{document}